\documentclass[9pt,twocolumn,twoside]{pnas-new}
\templatetype{pnasresearcharticle}

\usepackage[utf8]{inputenc}
\usepackage{amsmath,amssymb,graphicx,bm}

\usepackage{graphicx} 
\usepackage{subfigure} 
\usepackage{color}

\newcommand\beq{\begin{equation}}
\newcommand\eeq{\end{equation}}
\newcommand\beqa{\begin{eqnarray}}
\newcommand\eeqa{\end{eqnarray}}

\def\bal#1\eal{\begin{align}#1\end{align}}

\title{The Mpemba effect in spin glasses is a persistent memory effect}

\author[a,b]{Marco~Baity-Jesi}
\author[c]{Enrico~Calore}
\author[d,b]{Andres~Cruz}
\author[e,b]{Luis~Antonio~Fernandez}
\author[b]{Jos\'e~Miguel~Gil-Narvi{\'o}n}
\author[f,g,b]{Antonio~Gordillo-Guerrero}
\author[b,h]{David~I\~niguez}
\author[i]{Antonio~Lasanta}
\author[j,b]{Andrea~Maiorano}
\author[k]{Enzo~Marinari}
\author[e,b]{Victor~Martin-Mayor}
\author[b,d]{Javier~Moreno-Gordo}
\author[e,b]{Antonio~Mu\~noz~Sudupe}
\author[l]{Denis~Navarro}
\author[k,1]{Giorgio~Parisi}
\author[m,b,d]{Sergio~Perez-Gaviro}
\author[k]{Federico~Ricci-Tersenghi}
\author[n,g,b]{Juan~Jesus~Ruiz-Lorenzo}
\author[o]{Sebastiano~Fabio~Schifano}
\author[p,q,b]{Beatriz~Seoane}
\author[d,b]{Alfonso~Taranc{\'o}n}
\author[c]{Raffaele~Tripiccione}
\author[r,b,s,1]{David~Yllanes}

\affil[a]{Department of Chemistry, Columbia University, New York, NY 10027, USA}
\affil[b]{Instituto de Biocomputaci\'on y F\'{\i}sica de Sistemas Complejos (BIFI), 50018 Zaragoza, Spain}
\affil[c]{Dipartimento di Fisica e Scienze della Terra, Universit\`a di Ferrara e INFN, Sezione di Ferrara, I-44122  Ferrara, Italy}
\affil[d]{Departamento  de F\'\i{}sica Te\'orica, Universidad de Zaragoza, 50009 Zaragoza, Spain}
\affil[e]{Departamento  de F\'\i{}sica Te\'orica, Universidad Complutense, 28040 Madrid, Spain}
\affil[f]{Departamento de  Ingenier\'{\i}a El\'ectrica, Electr\'onica y Autom\'atica, U. de Extremadura, 10003, C\'aceres, Spain}
\affil[g]{Instituto de Computaci\'on Cient\'{\i}fica Avanzada (ICCAEx), Universidad de Extremadura, 06006 Badajoz, Spain}
\affil[h]{Fundaci\'on ARAID, Diputaci\'on General de Arag\'on, Zaragoza, Spain}
\affil[i]{Gregorio Mill\'an Institute of Fluid Dynamics,
Nanoscience and Industrial Mathematics,
Department of Materials Science and Engineering and Chemical Engineering,
Universidad Carlos III de Madrid, 28911 Legan\'es, Spain}
\affil[j]{Dipartimento di Fisica, Sapienza
  Universit\`a di Roma, 
  I-00185 Rome, Italy}
\affil[k]{Dipartimento di Fisica, Sapienza
  Universit\`a di Roma, INFN, Sezione di Roma 1, and CNR-Nanotec,
  I-00185 Rome, Italy}
\affil[l]{Departamento de Ingenier\'{\i}a, Electr\'onica y Comunicaciones and I3A, U. de Zaragoza, 50018 Zaragoza, Spain}
\affil[m]{Centro Universitario de la Defensa, Carretera de Huesca s/n, 50090 Zaragoza, Spain}
\affil[n]{Departamento de F\'{\i}sica, Universidad de Extremadura, 06006 Badajoz, Spain}
\affil[o]{Dipartimento di Matematica e Informatica, Universit\`a di Ferrara e INFN, Sezione di Ferrara, I-44122 Ferrara, Italy}
\affil[p]{Sorbonne Universit\'es, CNRS, IBPS, Laboratoire de Biologie Computationnelle et Quantitative - UMR 7238, 4 place Jussieu, 
75005 Paris, France}
\affil[q]{Sorbonne Universit\'es, Institut des Sciences du Calcul et des Donn\'ees, 4 place Jussieu, 
75005 Paris, France} 
\affil[r]{Department of Physics and Soft and Living Matter Program, Syracuse University, Syracuse, NY, 13244}
\affil[s]{Chan Zuckerberg Biohub, San Francisco, CA 94158}

\leadauthor{Baity-Jesi}

\date{\today}

\significancestatement{
The Mpemba effect, wherein an initially hotter system relaxes faster
when quenched to lower temperatures than an initially cooler system,
has attracted much attention. Paradoxically its very existence is a
hot topic. Using massive numerical simulations, we show unambiguously
that the Mpemba effect is present in the archetypical model system for
complex behavior, spin glasses. We find that the Mpemba effect in spin
glasses is due to the aging dynamics of the internal energy, as
controlled by the non-equilibrium spin-glass coherence length.
Interestingly, the effect is not present when the system remains all
the time in the paramagnetic phase. Therefore, the Mpemba effect
suggest itself as an effective probe for a glass transition,
potentially useful for other glass-forming systems.  
}

\authorcontributions{
A.L. and E.M. suggested undertaking
this project; L.A.F., A.L., V.M.-M. and J.J.R.-L. designed
the numerical experiment; the data analysis was performed by L.A.-F.,
A.L., V.M.-M., J.M.-G. and J.J.R.-L.; the physical interpretation of
results contained in the manuscript was provided by A.L. and V.M-M.,
with contributions from L.A.F., A.M., E.M.,J.M.-G., G.P., F.R.-T.,
J.J.R.-L. and D.Y.; D.I., S.F.S., A.T., and R.T. contributed Janus II
design; J.M.G.-N. and D.N. contributed Janus II simulation software;
M.B.-J., E.C., A.C., L.A.F., J.M.G.-N., A.G.-G., D.I., A.M., J.M.-G.,
A.M.S., S.P.-G., S.F.S., A.T., and R.T. contributed Janus II hardware
and software development; figures were contributed by M.B.-J., J.M.G,
L.A.F. and B.S.; the paper was written by M.B.-J., A.L., V.M.-M.,
J.M.-G., J.J.R.-L., F.R.-T., B.S. and D.Y.}

\authordeclaration{G.P. received funding from the Simons collaboration, and
Srikanth Sastry has received travel expenses to attend the Simons collaboration
annual meetings}

\correspondingauthor{\textsuperscript{1} To whom correspondence should be addressed. E-mail: giorgio.parisi@roma1.infn.it, david.yllanes@czbiohub.org}

\keywords{Spin glasses | memory effects | Mpemba effect | non-equilibrium physics}

\begin{abstract}
The Mpemba effect occurs when a hot system cools faster than an
initially colder one, when both are refrigerated in the same thermal
reservoir.  Using the custom built supercomputer Janus II, we study the
Mpemba effect in spin glasses and show that it is a
non-equilibrium process, governed by the coherence length $\xi$ of
the system. The effect occurs when the bath temperature lies in the
glassy phase, but it is not necessary for the thermal protocol to
cross the critical temperature. In fact, the Mpemba effect follows
from a strong relationship between the internal energy and $\xi$ that
turns out to be a sure-tell sign of being in the glassy phase.  Thus,
the Mpemba effect presents itself as an intriguing new avenue for
the experimental study of the coherence length in supercooled
liquids and other glass formers.
\end{abstract}

\dates{This manuscript was compiled on \today}
\doi{\url{www.pnas.org/cgi/doi/10.1073/pnas.XXXXXXXXXX}}

\begin{document}
\verticaladjustment{-2pt}
\newcommand{\Pe}{\ensuremath{\text{Pe}}}

\maketitle
\thispagestyle{firststyle}
\ifthenelse{\boolean{shortarticle}}{\ifthenelse{\boolean{singlecolumn}}{\abscontentformatted}{\abscontent}}{}
\renewcommand*{\thefootnote}{\fnsymbol{footnote}}
\dropcap{T}he Mpemba effect (ME) refers to the observation that the hotter
  of two identical beakers of water, put in contact with the same
  thermal reservoir, can cool faster under certain
  conditions~\cite{mpemba:69}.  The phenomenon is not specific to
  water, as it has been reported for nanotube
  resonators~\cite{greaney:11} and clathrate
  hydrates~\cite{ahn:16}. However, although records of the ME date as
  far back as Aristotle~\cite{aristotle-ross:81,jeng:06}, its very
  existence has been questioned~\cite{burridge:16}. 

  The ME is a non-equilibrium
process~\cite{lu:17,lasanta:17,vucelja:17} and 
when a complex system evolves out of equilibrium
its past history determines its fate. The simplest 
of these memory effects is, probably, the Kovacs
effect~\cite{kovacs:79}: after a temperature cycle, a glassy polymer
is left at the working temperature $T$ with a specific volume equal to
its equilibrium value.  Yet, the polymer is still out of equilibrium,
as evinced by a non-monotonic further time evolution of the specific volume.
Memory effects are ubiquitous and relevant in many systems of
technological and/or biological interest.  Important examples include
disordered materials~\cite{lahini:17}, polymers~\cite{struik:80},
amorphous solids~\cite{fiocco:14}, granular matter~\cite{prados:14}, biological
systems~\cite{kuersten:17}, batteries~\cite{sasaki:13} and, of course,
the disordered magnetic alloys known as spin glasses, which display
spectacular memory effects~\cite{jonason:98}.

Here, we show that spin glasses~\cite{young:98}  are an especially apt model
system to investigate the ME.  Using the Janus II
supercomputer~\cite{janus:14}, custom built for spin-glass simulations, we show
that the ME is indeed present in spin glasses and we clarify its origin: it is
a non-equilibrium memory effect, encoded in the glassy coherence length. 

\begin{figure}[!tb]
\includegraphics[width=\columnwidth,trim=20 40 20 35,clip]{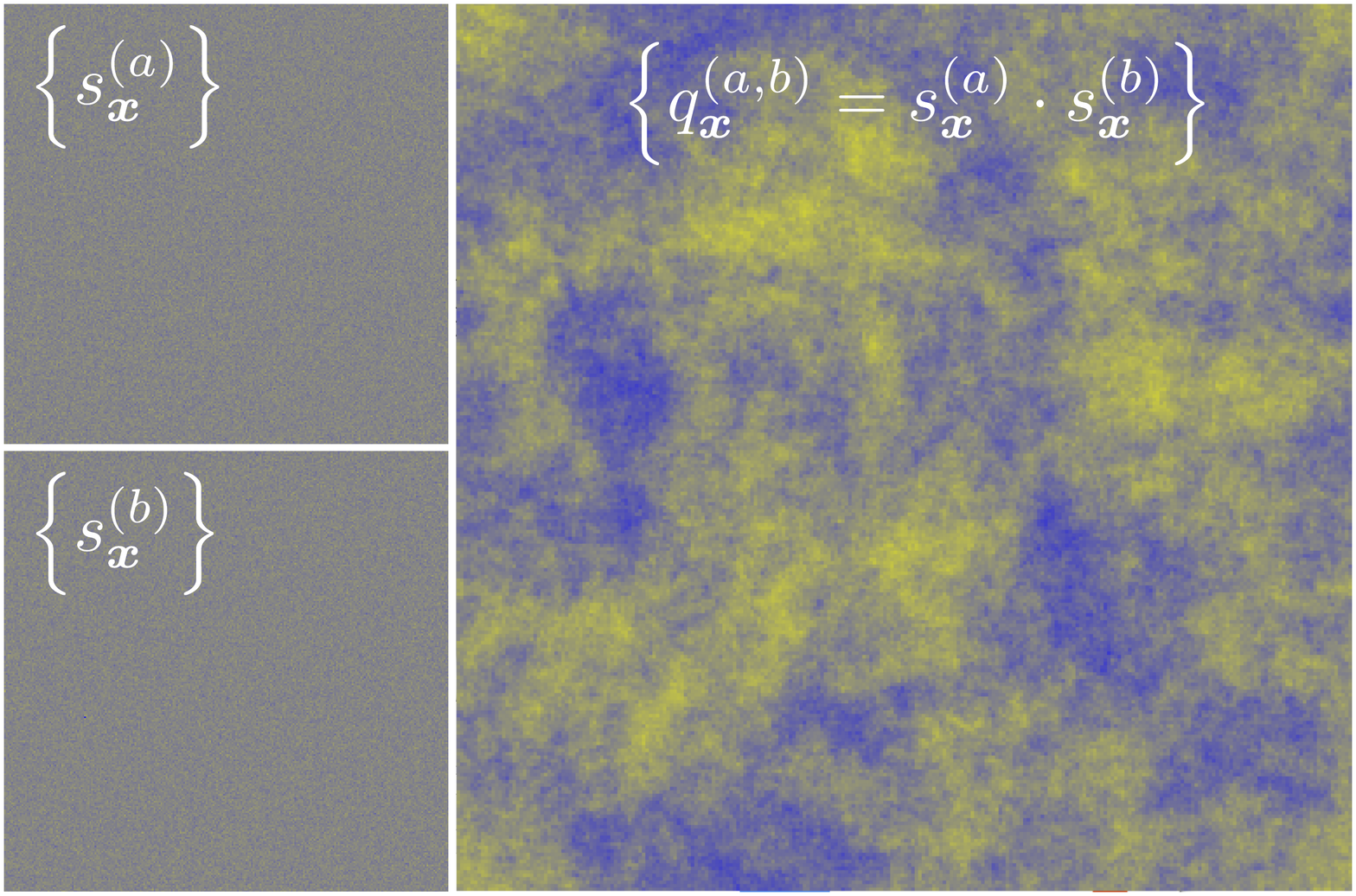}
\caption{\textbf{Spin-glass coherence length.} {\bf Top left:} A
  snapshot of a configuration $\{s_{\boldsymbol{x}}^{(a)}\}$, which
  has evolved for $t=2^{36}$ Monte Carlo steps at $T=0.7\approx
  0.64T_\text{c}$.  We show the average magnetization on the $xy$
  plane, averaging over $z$.  {\bf Bottom left:} Another configuration
  $\{s_{\boldsymbol{x}}^{(b)}\}$ of the same sample, prepared in the
  same way as $\{s_{\boldsymbol{x}}^{(a)}\}$.  No visible ordering is
  present in either configuration because the preferred pattern of the
  magnetic domains cannot be seen by eye ($s=1$ is plotted in yellow,
  and $-1$ in blue).  {\bf Right:} If one measures the overlap between
  the two configurations, see ~\eqref{eq:q-def} in
  Materials and Methods, and with the same color code used for the spins,
  the preferred pattern of the magnetic domains, of size $\xi$,
  becomes visible. A reliable method to extract $\xi$ from the overlap
  fields has been known for some time~\cite{janus:09b}, but only in
  2018 has it been possible to reach accuracies better than $1\%$
  in~$\xi$~\cite{janus:18}, thanks to the Janus~II
  computer~\cite{janus:14} and the use of many clones for the same
  sample (this strategy works only if the system size, which here is
  $L=160$, turns out to be much larger than $\xi$).  }
\label{fig:1}
\end{figure}

One major advantage of spin glasses as a model system is that, while
their  behavior is very complex, we now know 
that many of their major non-equilibrium processes
are ruled by the coherence length $\xi$ of the growing glassy
domains~\cite{marinari:96,berthier:02,janus:16,janus:17b}.
This includes, as we show in this paper, the ME. In the simplest protocol, a
sample initially at a high temperature is suddenly placed at
the working temperature $T$, lower that the critical temperature $T_\mathrm{c}$.
The system relaxes afterwards, but response functions such as
the magnetic susceptibility depend on time for as long as one has the
patience to wait (in analogy with living beings, glasses are said to
age~\cite{struik:80}). As aging proceeds, the size of the magnetic
domains, $\xi(t)$, perennially grows (yet, the magnetic
ordering is apparently random, see Fig.\ref{fig:1}). 

Only recently have we 
learned how to compute $\xi(t)$ reliably
from microscopic correlation
functions~\cite{janus:09b}. Interestingly enough, this
\emph{microscopic} $\xi(t)$ matches~\cite{janus:17b} the length scale
determined from \emph{macroscopic}
responses in experiments~\cite{guchhait:17}.

The time growth of $\xi(t)$ will be a crucial issue. In the paramagnetic phase,
$T>T_\mathrm{c}=1.102(3)$~\cite{janus:13}, $\xi$ grows up to its equilibrium
value, which can be very large close to $T_\mathrm{c}$. On the other hand, below
$T_\mathrm{c}$, $\xi(t)$ grows without bounds, but excruciatingly slowly.
Empirically, one finds that $t$ and $\xi(t)$ are related
through~\cite{janus:18,zhai:17}
\begin{equation}\label{eq:xi-slow}
\xi\propto t^{1/z(T)}\,,\quad z(T<T_\mathrm{c})\approx 9.6\,\frac{T_\mathrm{c}}{T}\,.
\end{equation}
The exponent $z(T)$, already very large near $T_\mathrm{c}$, becomes
even larger upon lowering $T$.  The very large value of $z(T)$
explains why supercomputers such as Janus~II, specifically designed to
simulate spin glasses~\cite{janus:14}, are necessary. Indeed, one has
to simulate the dynamics for a long time in order to see $\xi(t)$ vary
by a significant amount. The simulations discussed herein (see
Ref.~\cite{janus:18} and Materials and Methods) have $t$ varying by 11
orders of magnitude, for systems large enough to be representative of
the thermodynamic limit. On the other hand, for
$T>T_c$~\eqref{eq:xi-slow} holds only out-of-equilibrium, when $\xi$
has not yet approached its final equilibrium value. Under such
circumstances, $z~\approx 6$ has been computed
numerically~\cite{fernandez:15}.

\section*{Results}
\subsection*{The Mpemba effect in spin glasses}
The first step in a numerical study of the ME is identifying which
thermometric magnitude corresponds to the
off-equilibrium temperature that would be found experimentally.
In our case, this is the energy density (i.e., the instantaneous
energy per spin $E$, see Materials and Methods). $E$ is a natural
choice, because it is the observable conjugated with
temperature. Furthermore, \emph{in equilibrium}, there is a monotonically
increasing correspondence between $T$ and $E$.

\begin{figure}[tb!]
\centering
\includegraphics[width=\columnwidth]{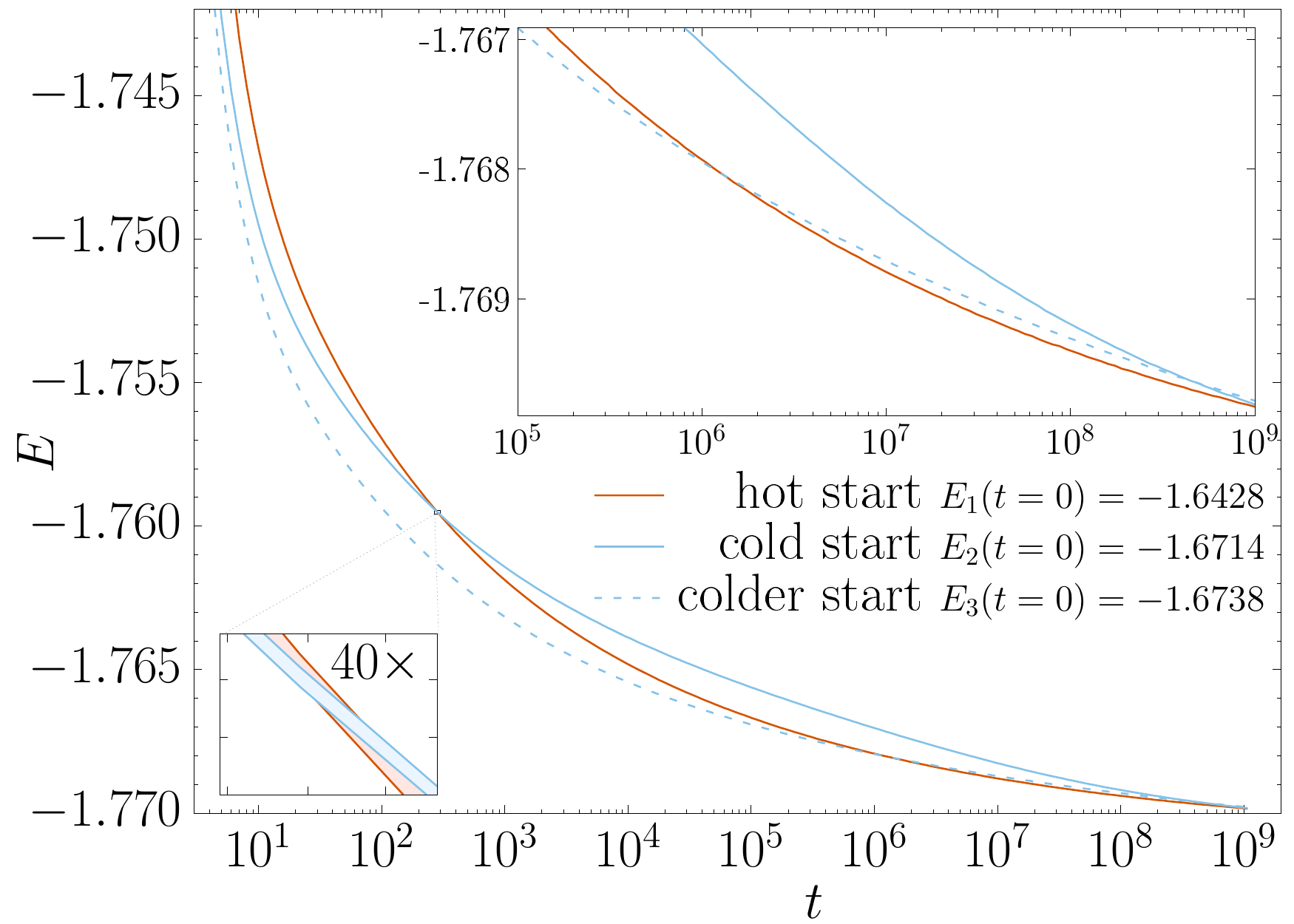}
\caption{{\bf Classical Mpemba protocol}. We show the time evolution
  of the energy of spin glass systems initially prepared at a higher
  temperature ($T=1.3$, red line) or at a lower temperature ($T=1.2$,
  blue lines), but always in the paramagnetic (high-temperature) phase
  ($T_\text{c}\approx1.10$).  In all three cases the systems are
  initially left to evolve out of equilibrium until they reach the
  internal energies shown in the figure key.  At $t=0$ all
  preparations are quenched, that is, put in contact with a thermal
  reservoir at temperature $T=0.7\approx0.64 T_\text{c}$.  As
  discussed in the text, the instantaneous internal energy is a
  measure of the (off-equilibrium) sample temperature. In agreement
  with the original Mpemba experiment~\cite{mpemba:69}, the system
  originally at the higher energy cools faster.  {\bf Bottom left
    inset:} Closeup of the first crossing between energy curves, showing
  the very small error bars, equal to the thickness of the lines
  (here and in all other figures, error bars represent one standard
  deviation; control variates, see~\cite{fernandez:09c} and Materials and Methods, 
  are used to improve accuracy).  {\bf Top right inset:}
  Closeup of the second crossing between energy curves.
} \label{fig:2}
\end{figure}

After the above considerations, we are ready to investigate the ME in
spin glasses (see Materials and Methods for details on the model and
observables).  Our first protocol strongly resembles the original
Mpemba experiment~\cite{mpemba:69}. We study the evolution of two
different non-equilibrium preparations for the same system: in
preparation~1 the system starts in a thermal bath at $T_1=1.3$ and is
left to evolve until it reaches an initial energy $E_1(t=0) \approx
-1.6428$, while in preparation~$2$ it is put in a bath at $T_2=1.2$
where it reaches a much lower initial energy $E_2(t=0) \approx
-1.6714$.\footnote{The configuration of the system before being
  exposed to either bath is random with a flat distribution (which is
  a typical $T=\infty$ configuration).}  Note that these states are
out of equilibrium: $T_1$ and $T_2$ only indicate the temperature of
the bath.  At time $t=0$, both systems are quenched, that is, they are
placed in contact with a thermal reservoir at $T=0.7 \approx 0.64
T_{\rm c}$. Fig.~\ref{fig:2} shows the evolution of the energies
(temperatures) after the quench: the hotter preparation (red) relaxes
to low energies faster than the colder preparation (solid blue).  This
is a perfect correspondence with the ME~\cite{mpemba:69}.

We also show with a blue dotted line the evolution of a third
preparation, starting again at the lower temperature $T=1.2$, but with a slightly
lower initial energy $E_3(t=0)\approx -1.6738$.  Even in this case the ME is present, but the crossing of the energy
curves takes place at a much longer time, 
even though the difference between the preparation energies $E_2$ and $E_3$ is of
only $0.15\%$. We need to find the controlling parameter.

\begin{figure}[tb!]
\centering
\includegraphics[width=\columnwidth]{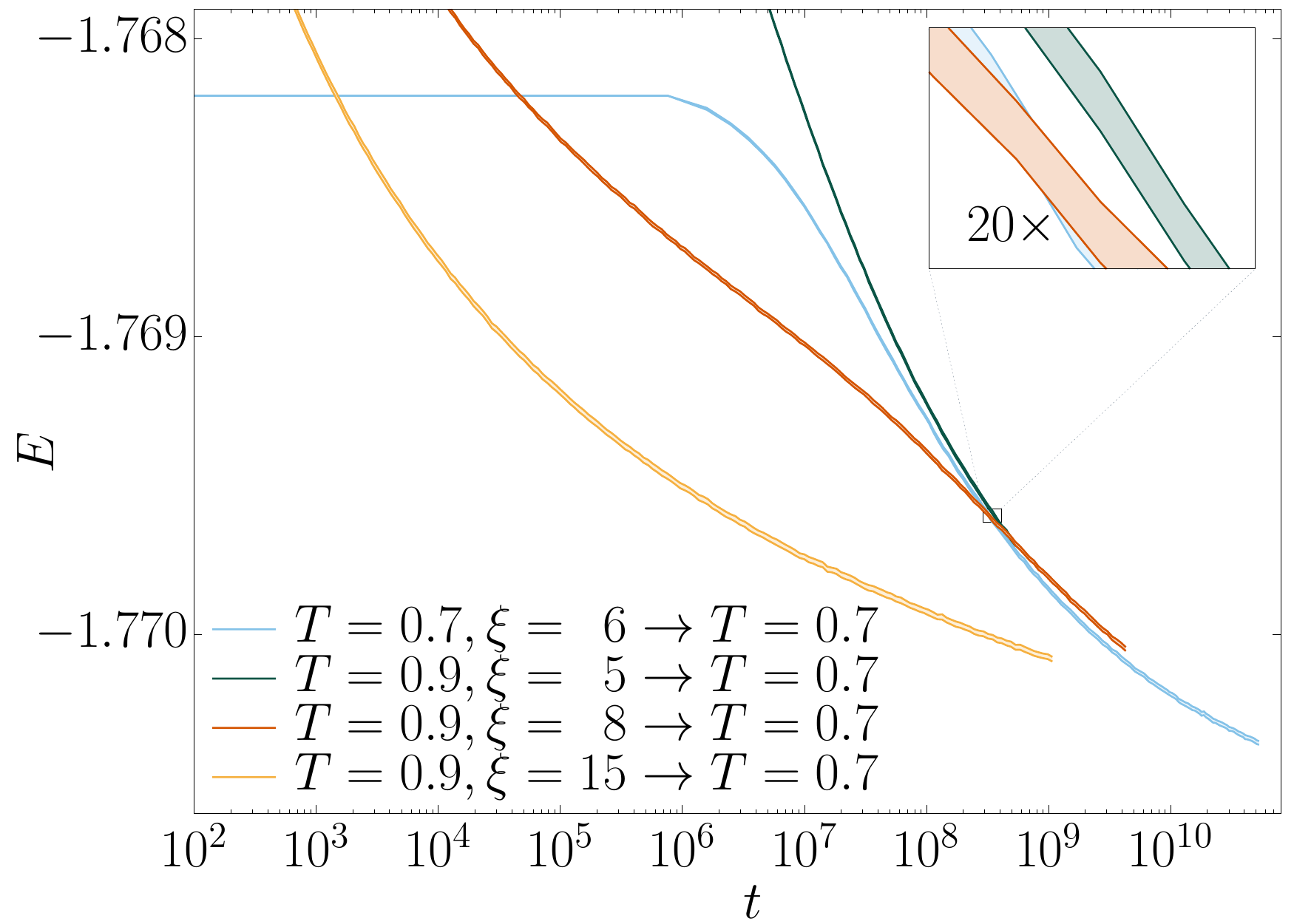}
\caption{{\bf Mpemba effect in the spin-glass phase.} As in
  Fig.~\ref{fig:2}, but all four initial preparations (see text for
  details) are now carried out in the spin-glass phase
  ($T<T_\text{c}$).  The preparation that cools faster is not the
  initially coldest one (blue curve), but the one with the largest
  initial coherence length (yellow curve). The last point in the
  isothermal relaxation at $T=0.7$ corresponds with the same time as
  the configurations shown in Fig.~\ref{fig:1}.  {\bf Inset:} A zoom
  of the \emph{second} crossing point between the curves for
  preparations $(T=0.9,\xi=8$; red curve) and $(T=0.7,\xi=6$; blue
  curve). This second crossing is not the ME. Rather, the ME arises at
  the \emph{first} crossing at $t\approx 5\times 10^4$. The second
  crossing disappears if one plots parametrically $E(t)$ as a function
  of $\xi(t)$. In other words, the description afforded by
  ~\eqref{eq:e-xi} is very accurate, but not exact.} \label{fig:3}
\end{figure}

Remembering our initial discussion of spin-glass dynamics a natural
candidate emerges: the coherence length $\xi(t)$.  Indeed, in terms of
$\xi$ our three starting conditions look very different. Our hot
preparation ($T_1=1.3, E_1=-1.6428$) had $\xi_1=12$, while our cold
preparations ($T_{2,3}=1.2$, $E_2=-1.6714$, $E_3=-1.6738$) had
$\xi_2=5$ and $\xi_3=8$, respectively. Therefore, upon quenching, the
initially hotter preparation is actually in a more advanced dynamical
state. In addition, preparations~2 and~3, which have almost the same
$E$, have very different starting $\xi$, which, from
Eq.~(\ref{eq:xi-slow}), can explain the differing relaxation times.

We arrive, then, at our working hypothesis: out of equilibrium, our system
is not adequately labeled by the temperature $T$ of the thermal bath alone,
or even by $T$ plus the instantaneous internal energy $E$. 
Instead, $\xi$ emerges as the hidden parameter that will
rationalize the results.

Notice that, according to this idea, crossing $T_\text{c}$
should not be necessary in order to find an ME.
The only required ingredient would be starting points A and B
with $T_\text{A}>T_\text{B}$ and $\xi_\text{A}>\xi_\text{B}$.
We test this hypothesis in Figure~\ref{fig:3},
where we work solely in the low-temperature phase.
Preparation~1 starts at $T=0.7$ until it reaches $\xi_1=6$,
while preparations 2--4 evolve at $T=0.9$ until they 
reach $\xi=5,8,15$, respectively. All samples are then quenched 
to $T=0.7$ and we start measuring (i.e., we set $t=0$).
In accordance with our previous discussion, the cooling rate
is not controlled by the preparation temperature, but by the 
starting coherence length. Preparations starting at $T=0.9$  but 
with $\xi(t=0)$ greater than $\xi_1=6$ cool faster 
than preparation 1.

\subsection*{The energy-coherence length phase diagram}\label{sect:phase-diagram}

In order to make these observations more quantitative, we need to
explore the relationship between $E$ and $\xi$ in an off-equilibrium
spin glass. Fortunately, we have heuristic arguments~\cite{parisi:88}
and numerical results~\cite{marinari:96,janus:09b} suggesting that
\begin{equation}\label{eq:e-xi}
E(t)=E_\infty(T)+\frac{E_1}{\xi^{D_\ell}(t)}+\ldots\,.
\end{equation}
The dots stand for scaling corrections, subdominant for large $\xi$,
and $D_\ell\approx2.5$~\cite{boettcher:05} is the lower critical
dimension (the phase transition occurs only for spatial dimension
$D>D_\ell$). We note that the heuristic derivation of
~\eqref{eq:e-xi}, recall Ref. ~\cite{parisi:88}, makes sense only in
the spin-glass phase, $T<T_{\mathrm{c}}$. We shall test below and in
our study of the inverse Mpemba Effect whether or not ~\eqref{eq:e-xi}
applies in the spin-glass phase and in the paramagnetic phase.

\begin{figure}[t!]
\centering
\includegraphics[width=\columnwidth]{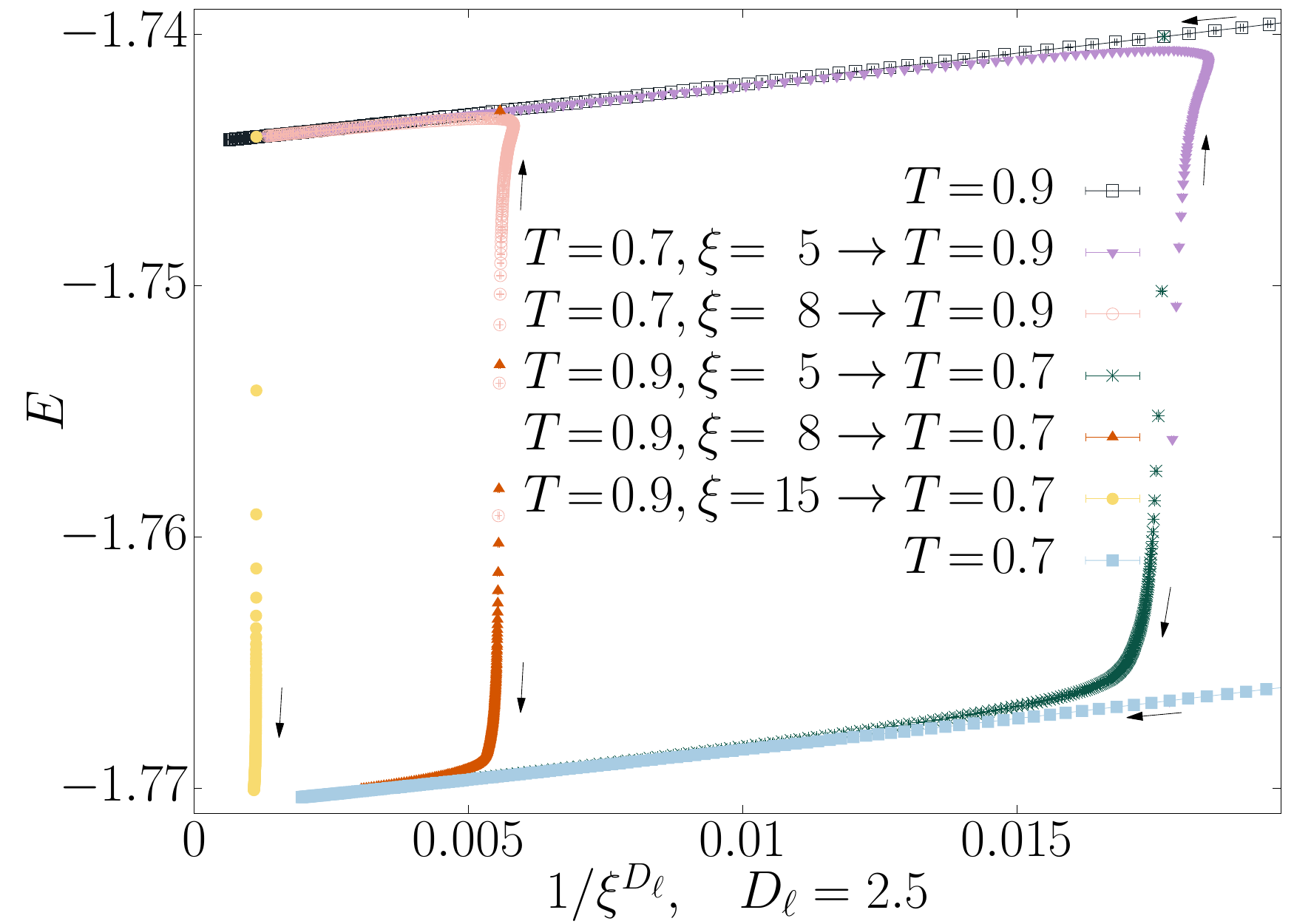}
\caption{{\bf Relationship between the energy density $E$ and the
    coherence length $\xi$}. As suggested by ~\eqref{eq:e-xi}, for
  isothermal relaxations (in the plot $T=0.7$ and $0.9$, depicted with
  continuous lines) $E$ is an essentially linear function of
  $1/\xi^{2.5}$ (at least for the plotted range of $\xi>4.8$).
  Furthermore, the dependence  of the slope on temperature is
  marginal.  Temperature-varying protocols are seen to be essentially
  vertical moves between the straight lines corresponding to
  isothermal relaxations at the initial and final temperatures. These
  vertical moves are very quick initial transients, in which (in moves
  to higher temperatures only), $\xi$ slightly decreases and then
  increases again.  } \label{fig:4}
\end{figure}

As Fig.~\ref{fig:4} shows, the ME in spin glasses follows from the
combination of two simple ideas: 
\begin{enumerate}
\item When a quick temperature change takes place, $\xi$ [which is a
  slow variable, recall ~\eqref{eq:xi-slow}] has no time to adjust
  and remains basically unchanged.
\item Setting quick transients aside, the energy density follows in all cases
~\eqref{eq:e-xi}. 
\end{enumerate}
Both of these points are slight oversimplifications
(the analysis of the deviations will be performed elsewhere)
but their combination results in a very simple
explanation of the ME. We first note that, when temperature
is fixed, the energy density is an (almost) linear function of
$1/\xi^{2.5}$. Furthermore, the slope $E_1$ varies by less than
$4\%$ in the range we explore. It follows that the
natural phase diagram to discuss the ME is the
($E$,$1/\xi^{2.5}$) plane.  Indeed, isothermal relaxations are
represented in the ($E$,$1/\xi^{2.5}$) plane as a set of (almost)
parallel, (almost) straight lines. To an excellent first
approximation, the effects of temperature changes can be depicted as
almost instantaneous vertical moves between the parallel straight
lines that correspond to the initial and final temperatures. It then follows
that, in the spin-glass phase at least,
the larger the starting $\xi$, the faster the cooling, irrespective
of the initial preparation protocol, exactly as
observed in Figs.~\ref{fig:2} and~\ref{fig:3}.

\subsection*{Inverse Mpemba Effect}\label{sect:IME}

In Refs.~\cite{lasanta:17,lu:17} an inverse thermal protocol was
suggested: the bath temperature was chosen to be higher than the one
of the starting condition.  Under such circumstances, it was observed
that the coolest of two initial preparations could heat faster. This
effect was named \emph{inverse} ME (IME).

As we have seen in the discussion of the energy-coherence length phase
diagram, heating and cooling protocols are quite symmetrical
\emph{provided that both the final and initial bath temperatures lie
  below} $T_{\mathrm{c}}$. It is therefore more challenging to see
whether or not the IME survives crossing $T_{\mathrm{c}}$. The
question is non-trivial, because ~\eqref{eq:e-xi} does not hold in the
paramagnetic phase.

To verify this, we reverse our protocol, choosing starting
temperatures $T<T_\mathrm{c}$ in the spin-glass phase, and warm up the
samples to a temperature well above $T_\mathrm{c}$.  Specifically, we
use three starting conditions for the warm-up experiment.  In
protocol 1, we choose $T=0.7$ and $\xi=2.5$; in protocol 2, we choose
$T=0.7$ and $\xi=11.7$; in protocol 3, we choose $T=0.8$ and
$\xi=15.8$.  In the three cases, we switch instantly the thermal bath
to $T=1.4$ [where the asymptotic equilibrium value is $\xi=8.95(5)$],
and follow the relaxation of coherence length and energy.

\begin{figure}[t!]
\centering
\includegraphics[width=\columnwidth]{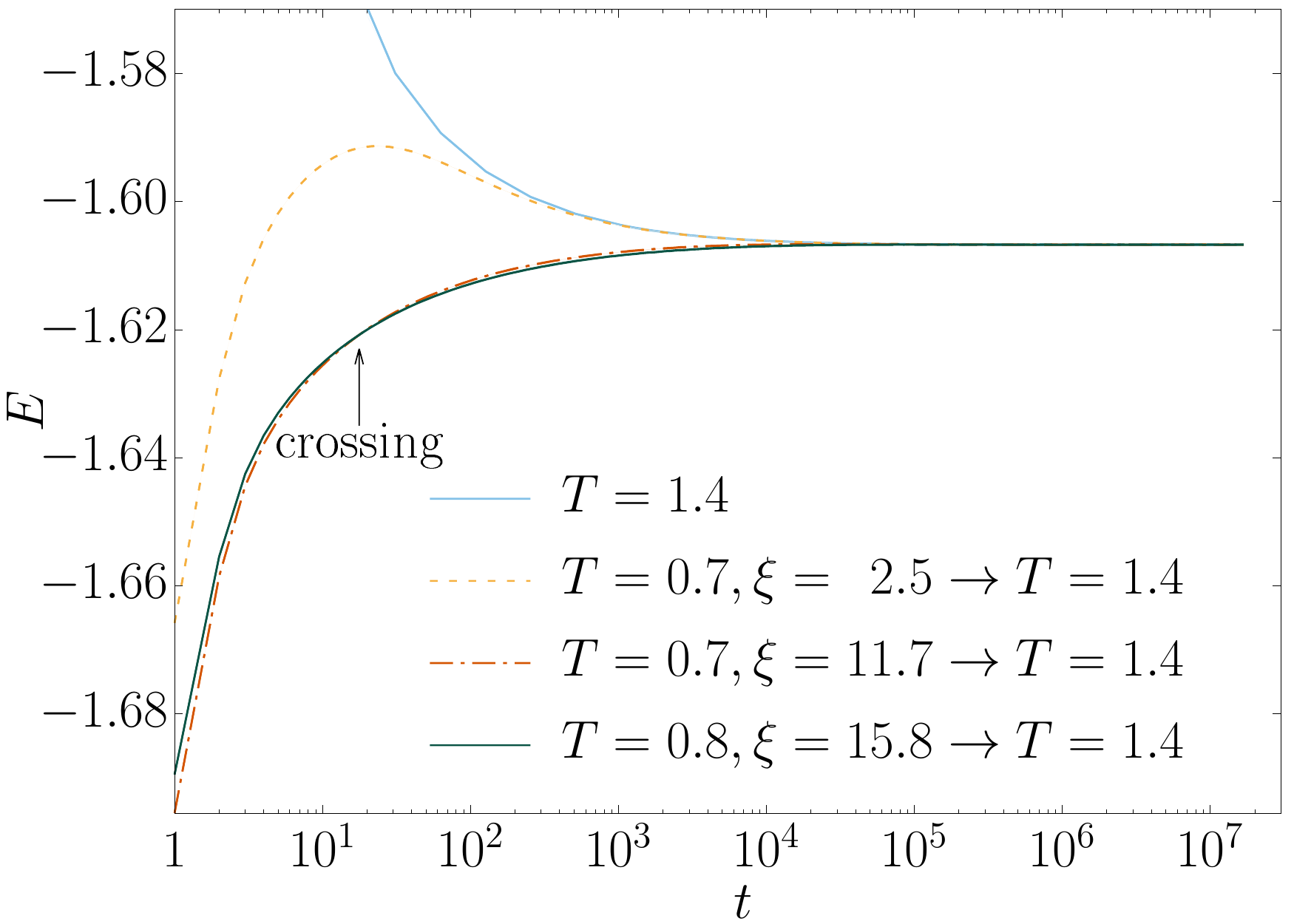}
\caption{\textbf{A tiny inverse Mpemba effect.}  Time evolution of the
  energy, for the three different preparations described in the main
  text, compared with a quench from $T=\infty$ to $T=1.4$ (top curve).
  In the three cases, the initial temperature is in the spin glass
  phase, and the final temperature is $T=1.4>T_\mathrm{c}$. A very
  small ME is found at the time pointed by the arrow, only when
  warming up samples with similar starting energy.} \label{fig:I2}
\end{figure}

Fig.~\ref{fig:I2} depicts the evolution of the energy in the warm-up
experiments. If one compares protocol 1 with protocol 3, the ME is
clearly absent.  Instead, a tiny ME is present when comparing
protocols~2 and~3, which have a more similar starting energy, and a
coherence length that manifests a better thermalization in the spin
glass phase: the curve that is initially hotter becomes colder after
$\sim20$ iterations.

\begin{figure}[t!]
\centering
\includegraphics[width=\columnwidth]{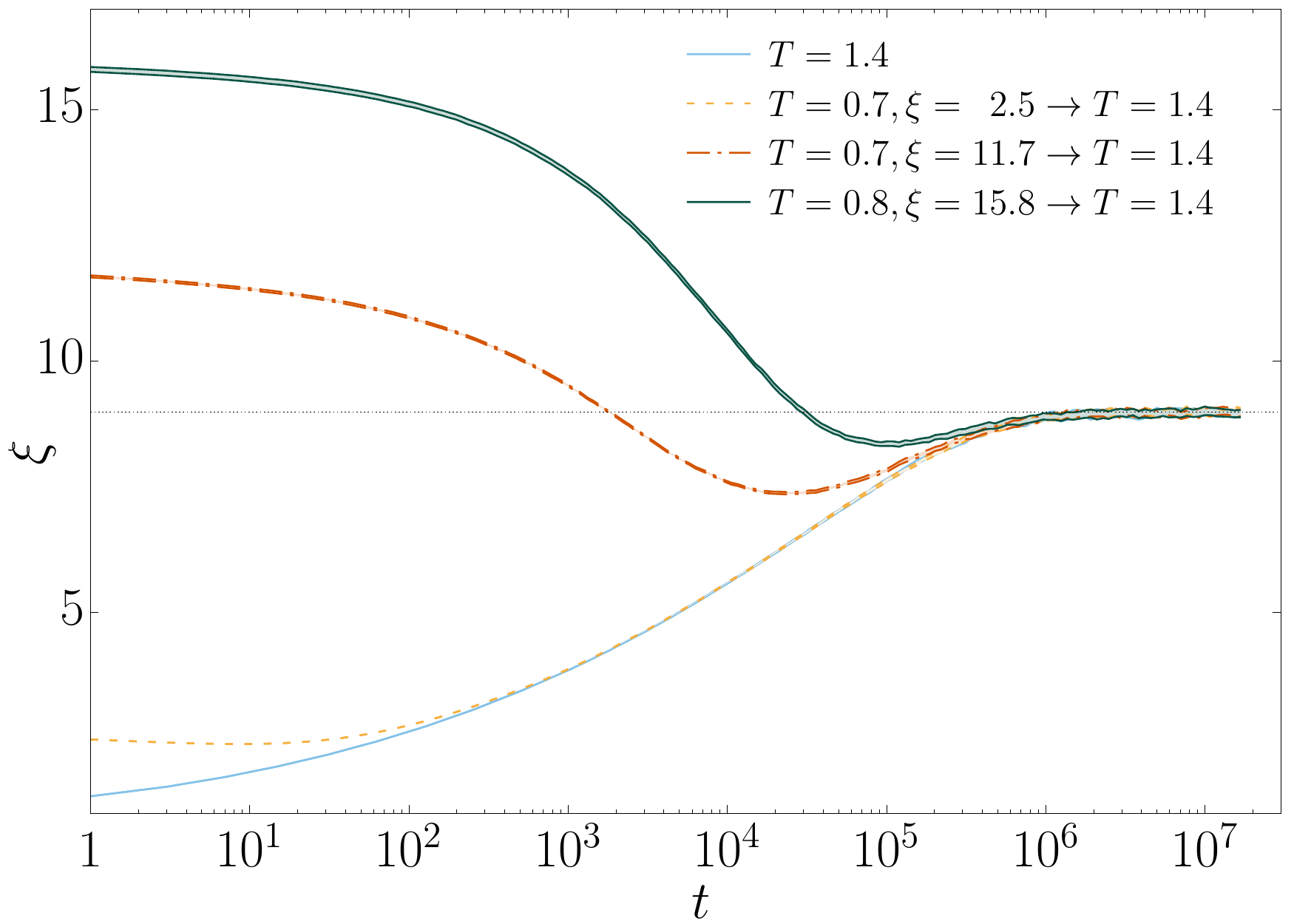}
\caption{\textbf{Coherence length: Undershooting and convergence to a master curve}.
Coherence lengths $\xi$ of the experiments described in Fig.~\ref{fig:I2}. 
The time evolution of $\xi$ tends to converge towards the curve corresponding
to a quench from $T=\infty$ to $T=1.4$ (bottom curve),
giving rise to an undershoot of $\xi$ when its initial value is higher than
the equilibrium $\xi$ at $T=1.4$.} \label{fig:I1}
\end{figure}

In Fig.~\ref{fig:I1}, we show the evolution of $\xi$ during the three
warm-up protocols.  Visibly, ~\eqref{eq:xi-slow}, valid only under
$T_\mathrm{c}$, does not hold.  Furthermore, for all of the starting
conditions the curves tend to approach a master curve represented by
the instant quench from $T=\infty$ to $T=1.4$, generating an undershoot
in the time evolution of the observable.  An analogous independence of
the initial conditions has been recently observed in the paramagnetic
phase of a two-dimensional spin glass
\cite{fernandez:18a,fernandez:18b}.

By comparing Figs.~\ref{fig:I2} and~\ref{fig:I1}, one can see that
energy and coherence length converge to their equilibrium value at
very different times, and that an undershoot on $E(t)$ is not
corresponded with an undershoot of $\xi(t)$ (and vice versa).  This
decoupling between the two observables is necessary for the ME to take
place, since the time scale of the crossing between protocols~2 and~3
does not correspond to any significative change in coherence lengths.
The non-monotonic behavior of $E(t)$ and $\xi(t)$ will be the  subject
of further investigation.

\section*{Discussion}
In summary, we have shown that the Mpemba effect (ME) is present
in spin glasses, where it appears as an intrinsically
non-equilibrium process, ruled by the spin-glass coherence
length $\xi$. Although encoding the history of a complex
system by a single number is an oversimplification,
we have shown that the approximate
description afforded by ~\eqref{eq:e-xi} is accurate enough to
explain the ME. 

Our results explain how the most natural experimental setup (prepare
two identical systems at $T_1,T_2>T_\text{c}$ with an identical
protocol, then quench them) can fail to see the
effect~\cite{burridge:16}.  Indeed, for spin glasses at least, a
different starting $\xi$ is required. Above $T_\text{c}$, where the
growth rate of $\xi$ does not depend on $T$, this means letting the
hotter preparation evolve for longer time at the initial temperature
before the quench.  On the other hand, if we prepare the systems at
$T_1,T_2<T_\text{c}$ and then bring them to an even lower temperature,
the effect is enhanced, because in the $T<T_\text{c}$ phase the growth
of $\xi$ is slower for lower temperatures.  Therefore, in the
low-temperature phase the ME can be found even with identical
preparation times for the hot and cold preparations.  Finally, we have
investigated the Inverse ME (IME). If both the initial and the final temperatures
lie below $T_{\mathrm{c}}$ the system behaves in a symmetrical way
under cooling or heating (see Fig.~\ref{fig:4}). On the other hand, if
we start in the spin glass phase and end in the paramagnetic phase,
the IME is strongly suppressed because ~\eqref{eq:e-xi} is not
valid for $T>T_{\mathrm{c}}$.

The ME is peculiar among the many memory effects present in
spin glasses. Indeed, this phenomenon can be studied through quantities such as
the energy density, which are measured at just one time scale (rather than
the usual two times~\cite{young:98,jonason:98,janus:17b}). However, our setup
poses an experimental challenge, because we are not aware of any
measurement of the non-equilibrium temperature associated with the
magnetic degrees of freedom. Perhaps one could adapt the
strategy of Ref.~\cite{grigera:99}, connecting dielectric susceptibility and
polarization noise in glycerol, to measurements of high-frequency electrical
noise in spin glasses~\cite{israeloff:89}.

An easier experimental avenue is suggested by our study of the  inverse Mpemba
effect, where the preparations are heated, rather than cooled. In this case,
while the response of the energy is very small, the process is accompanied by a
dramatic memory effect in the coherence length. This quantity has a
non-monotonic time evolution upon heating from the spin-glass to the
paramagnetic phase, before converging to the master (isothermal) curve and is
measurable with current experimental techniques.

Our investigation of the Mpemba effect offers as well a new
perspective into an important problem, namely the study of the glassy
coherence length in supercooled liquids and other glass
formers~\cite{cavagna:09}. Indeed, the identification of the
right correlation function to study experimentally (or
numerically) is still an open problem. Spin glasses are unique in the
general context of the glass transition, in both senses: we know which
correlation functions should be computed
microscopically~\cite{edwards:75,edwards:76}, while accurate
experimental determinations of the coherence length have been
obtained~\cite{guchhait:17}. The fact that we have such a strong
command over the spin-glass coherence length has allowed us to test
to very high accuracy its relationship with an extremely simple
quantity such as the energy, see ~\eqref{eq:e-xi}. Furthermore, we
have shown that this tight relationship between the coherence length
and the energy holds only in the spin-glass phase. Yet, the energy is
a local, single-time quantity just as the specific volume measured in
the Kovacs effect in polymers. Therefore, ~\eqref{eq:e-xi} suggests
an intriguing and far reaching alternative: rather than considering
two-time correlation functions of ever-increasing complexity, it
might be worth seeking high-accuracy experimental measurements of
single-time quantities such as the specific volume.  Furthermore, we find
that the Mpemba effect is tiny, almost invisible, in the paramagnetic
phase. Thus, finding a sizeable Mpemba-like behavior for quantities
such as the specific volume may be a crucial step forward in the
experimental identification of a glassy coherence length in a
low-temperature phase.

\matmethods{\subsection*{Model and observables}
We consider  Metropolis dynamics for the Edwards-Anderson model in
a cubic lattice of linear size $L=160$, with nearest-neighbor
interactions and periodic boundary conditions:
\begin{equation}
{\cal H}=-\sum_{\langle \boldsymbol{x}, \boldsymbol{y}\rangle } J_{\boldsymbol{x
},\boldsymbol{y}} s_{\boldsymbol{x}}\, s_{\boldsymbol{y}}\,.\label{eq:EA-H}
\end{equation}
The spins $s_{\boldsymbol{x}}\!=\!\pm1$ are placed on the lattice
nodes, $\boldsymbol{x}$. The couplings
$J_{\boldsymbol{x},\boldsymbol{y}}\!=\!\pm 1$, which join nearest
neighbors only, are chosen randomly with $50\%$ probability and are
quenched variables. For each choice of the couplings (one ``sample''),
we simulate 256 independent copies of the system (clones).  We denote
by $\langle\cdots\rangle_J$ the average over the thermal noise
(i.e. average over results for the 256 clones corresponding to a
single sample), the \emph{subsequent} average over our 16 samples is
indicated by square brackets $[\, (\cdot \cdot \cdot)\,]$. The
lattice size $L=160$ is large enough to effectively represent the
thermodynamic limit (for this, and other details, see
Ref.~\cite{janus:18}). The model described by \eqref{eq:EA-H}
undergoes a spin-glass transition at
$T_\mathrm{c}=1.102(3)$~\cite{janus:13}. A single Metropolis sweep
roughly corresponds to one picosecond of physical time. Therefore, the
time in our simulations varies from one picosecond to a tenth
of a second.

We compute the spin-glass coherence length from the
decay of the  microscopic correlation function
\begin{equation} \label{eq:C4-def}
C_4(\boldsymbol{r},t)=\frac{1}{L^3}\sum_{\boldsymbol{x}}\, [\,\langle 
q_{\boldsymbol{x}}^{(a,b)} q_{\boldsymbol{x}+\boldsymbol{r}}^{(a,b)}\rangle_J \,]\,.
\end{equation}
where the overlap-field is computed as
\begin{equation}\label{eq:q-def}
q_{\boldsymbol{x}}^{(a,b)}=s_{\boldsymbol{x}}^{(a)} s_{\boldsymbol{x}}^{(b)}\,.
\end{equation}
In the above expression, indices $a$ and $b$ label \emph{different}
clones (of course, we average over the $256\times255/2$ possible
choices for the pair of clones). Technically speaking, the spin-glass
coherence length $\xi$ employed in this work corresponds to the
$\xi_{1,2}$ integral determination (we refer the reader to the
literature for details~\cite{janus:09b,janus:18}).

\begin{figure}[t]
\centering
\includegraphics[width=\columnwidth]{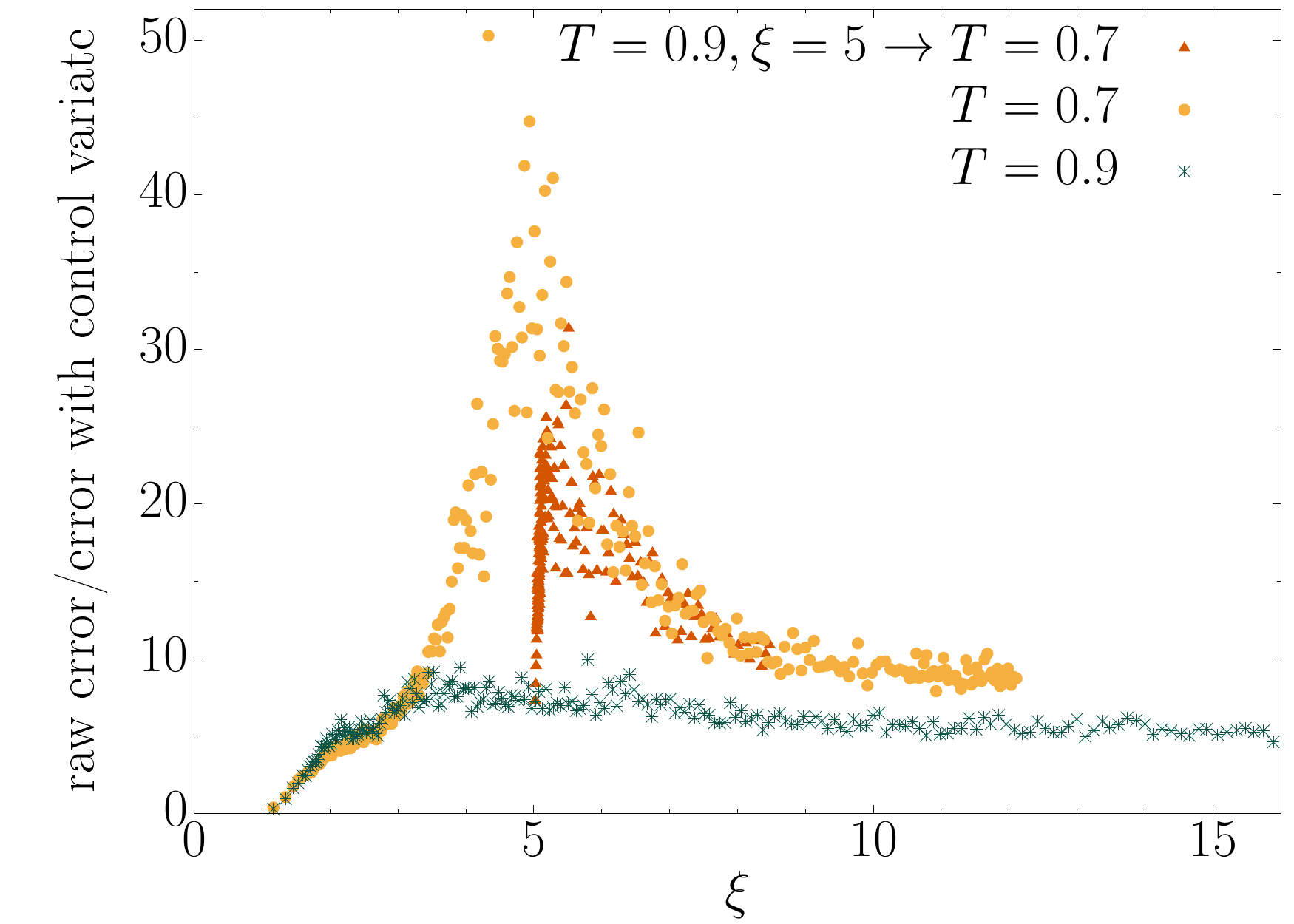}
\caption{{\bf Improving the accuracy with control variates}. The
  figure shows the ratio of statistical errors, as a function of
  $\xi(t)$, for the naive [~\eqref{eq:e-naive}] and improved
  [~\eqref{eq:e-cv}] estimates of the energy density. The data
  shown correspond to three different relaxations. Two of them are
  isothermal relaxations starting from $\xi=0$ at $t=0$. The third
  relaxation corresponds to the preparation starting at
  $(T=0.9,\xi=5)$ which is quenched to $T=0.7$ at $t=0$ (i.e. the green curves in Figs.~\ref{fig:3} and~\ref{fig:4}). The error
  reduction is largest for the isothermal relaxation at $T=0.7$ and
  $2^{19}\leq t\leq 2^{21}$, of course (after all, this is the
  temperature and time region defining the control variate), but the
  error reduction is also very significant at other times and
  temperatures.}\label{fig:5}
\end{figure}

\subsection*{The energy density and its control variate.}
The naive determination of the internal energy density at time $t$
($t$ is the time elapsed after the initial preparation, as measured in Metropolis full-lattice sweeps) is
\begin{equation}\label{eq:e-naive}
E^\mathrm{naive}(t)= \frac{1}{L^3} \overline{\langle {\cal H}(t)\rangle_J} \,.
\end{equation}
This estimate, which is perfectly correct, can be made far more
accurate by using a suitable control variate~\cite{fernandez:09c}. 

Before going on, it should be clear that $E^\mathrm{naive}(t)$ is a
random variable: it is the estimate of the thermal and disorder
average~\emph{as obtained from our numerical data}. Of course,
the~\emph{exact} mean values, obtained by averaging over the thermal noise and 
the coupling realizations with infinite statistics, are not random
variables.

Specifically, we improve $E^\mathrm{naive}(t)$ by subtracting from it
another random variable, named the control variate, extremely
correlated with it. The expectation value of the control variate
vanishes, hence it does not change the (inaccessible with finite
statistics) exact expectation value of
$E^\mathrm{naive}(t)$. Furthermore, we choose a control variate which
is independent of temperature and time. So, the overall effect of the
control variate is twofold: (i) a uniform (random) vertical shift for
all data in Figs.~\ref{fig:2}--\ref{fig:I2}
 and (ii) a dramatic reduction of the error bars.

Our control variate is obtained as follows. For each of our 16 samples, we take a fully
random initial configuration, place it suddenly at $T=0.7$ (this is our $t=0$),
simulate it at constant temperature and, finally, compute 
\begin{equation}
{\cal E}_J=\frac{1}{L^3\, N_S} \sum_{t\in S} \langle {\cal H}(t)\rangle_J\,, 
\end{equation}
where $S$ is the set of the $N_S$ times of the form $t$=(integer part
of) $2^{n/8}$, with $n$ integer and $2^{19}\leq t\leq 2^{21}$. Besides, we
obtain a high-accuracy estimate of the same quantity, ${\cal E}^\mathrm{accurate}
$ by averaging over the results of a
targeted simulation with 160000 samples and 2 clones (i.e. the energy
is averaged over 320000 systems).  The final estimates of the internal
energy shown in Figs.~\ref{fig:2}--\ref{fig:I2}
are
\begin{equation}\label{eq:e-cv}
E(t)=E^\mathrm{naive}(t)\ -\  [\, {\cal E}_J -{\cal E}^\mathrm{accurate}\,]\,.
\end{equation}
The benefits of using the control variate are obvious from Fig.~\ref{fig:5}.}
\showmatmethods

\acknow{
We strongly thank Raymond Orbach, Srikanth Sastry
and Marija Vucelja for very interesting comments. This work was
partially supported by Ministerio de Econom\'ia, Industria y
Competitividad (MINECO), Spain, through Grants No.  FIS2013-42840-P,
No. MTM2014-56948-C2-2-P, No. FIS2015-65078-C2, No. FIS2016-76359-P,
No.  TEC2016-78358-R and No. MTM2017-84446-C2-2-R (Grants also partly
funded by FEDER); by the Junta de Extremadura (Spain) through Grant
Nos.  GRU18079 and IB16013 (partially funded by FEDER) and by the
DGA-FSE (Diputaci\'on General de Arag\'on -- Fondo Social Europeo).
This project has received funding from the European Research Council
(ERC) under the European Union's Horizon 2020 research and innovation
program (Grant Nos. 694925 and 723955 - GlassUniversality).
D.Y. acknowledges support by the Soft and Living Matter Program at
Syracuse University.  EM and DY thank the KITP and its ``Memory
Formation in Matter'' programme, where contributions to the present
work were developed, and acknowledge support from the National Science
Foundation under Grant No. from NSF-PHY-1748958.}
\showacknow


\end{document}